\documentclass[11pt]{article}

\usepackage{epsfig}

\newcommand{\sect}[1]{\setcounter{equation}{0}\section{#1}}
\renewcommand{\theequation}{\arabic{section}.\arabic{equation}}

\def\be{\begin{equation}}
\def\ee{\end{equation}}
\def\ba{\begin{eqnarray}}
\def\ea{\end{eqnarray}}

\title{{\bf Brane-World Black Holes}}
\author{A. Chamblin\thanks{email: H.A. Chamblin@damtp.cam.ac.uk},
S.W. Hawking\thanks{email: S.W.Hawking@damtp.cam.ac.uk} and
H.S. Reall\thanks{email: H.S.Reall@damtp.cam.ac.uk}
 \\ DAMTP \\ University of Cambridge \\ Silver Street,
Cambridge CB3 9EW, United Kingdom. \\ \\ Preprint DAMTP-1999-133}

\date{29 September 1999}

\begin{document}

\maketitle

\begin{abstract}

Gravitational collapse of matter trapped on a brane will produce a
black hole on the brane. We discuss such black holes in the models of
Randall and Sundrum where our universe is viewed as a domain wall in
five dimensional anti-de Sitter space. We present evidence that a
non-rotating uncharged black hole on the domain wall is 
described by a ``black cigar'' solution in five dimensions.

\end{abstract}

\sect{Introduction}

There has been much recent interest in the idea that our universe may
be a brane embedded in some higher dimensional space. It has been
shown that the hierarchy problem can be solved if the higher
dimensional Planck scale is low and the extra dimensions large
\cite{large1,large2}. An alternative solution, proposed by Randall and
Sundrum (RS), assumes that our universe is a
negative tension domain wall separated from a positive tension wall by
a slab of anti-de Sitter (AdS) space \cite{RS1}. This does not require
a large extra dimension: the hierarchy problem is solved by the
special properties of AdS. The drawback with this model is the
necessity of a negative tension object. 

In further work \cite{RS2}, RS suggested that it is
possible to have an {\it infinite} extra dimension. In this model,
we live on a positive tension domain wall inside anti-de Sitter space. There
is a bound state of the graviton confined to the wall as well as a
continuum of Kaluza-Klein (KK) states. For non-relativistic processes
on the wall, the bound state dominates over the KK states to give an
inverse square law if the AdS radius is sufficiently small. It
appears therefore that four dimensional gravity is recovered on the
domain wall. This conclusion was based on perturbative
calculations for zero thickness walls. Supergravity 
domain walls of finite thickness have
recently been considered \cite{cvetic,townsend,gubser} and a non-perturbative
proof that the bound state exists for such walls was given in
\cite{ppwave}. It is important to examine other non-perturbative 
gravitational effects in this scenario to see whether the predictions 
of four dimensional general relativity are recovered on the domain wall.

If matter trapped on a brane undergoes gravitational collapse then a
black hole will form. Such a black hole will have a horizon that
extends into the dimensions transverse to the brane: it will be a
higher dimensional object. Phenomenological properties of such black
holes have been discussed in \cite{submmbh} for models with 
large extra dimensions. In this paper we discuss black holes in the
RS models. A natural candidate for such a hole is the Schwarzschild-AdS
solution, describing a black hole localized in the fifth dimension. We
show in the Appendix that it is not possible to intersect 
such a hole with a {\it vacuum} domain wall so it is unlikely 
that it could be the final state of gravitational collapse on the
brane. A second possibility is
that what looks like a black hole on the brane is actually a black
string in the higher dimensional space. We give a simple solution
describing such a string. The induced metric on the domain wall is
simply Schwarzschild, as it has to be if four dimensional general
relativity (and therefore Birkhoff's theorem) are recovered on the
wall. This means that the usual astrophysical properties of black holes
(e.g. perihelion precession, light bending etc.) are recovered in this
scenario.

We find that the AdS horizon is singular for 
this black string solution. This is
signalled by scalar curvature invariants diverging if one approaches
the horizon along the axis of the string. If one approaches
the horizon in a different direction then no scalar curvature
invariant diverges. However, in a frame parallelly propagated 
along a timelike geodesic, some curvature components {\it do} diverge. 
Furthermore, the black string is unstable near the AdS
horizon - this is the Gregory-Laflamme instability
\cite{greg}. However, the solution is stable far from the AdS
horizon. We will argue that our solution evolves to a ``black
cigar'' solution describing an object that looks like the black string
far from the AdS horizon (so the metric on the domain wall is
Schwarzschild) but has a horizon that closes off before reaching the
AdS horizon. In fact, we conjecture that this black cigar
solution is the unique stable vacuum solution in five dimensions
which describes the endpoint of gravitational collapse on the
brane.  We suspect that the AdS horizon will be non-singular for
the cigar solution. 

\sect{The Randall-Sundrum models}

\label{sec:rs}

\begin{figure}
\centerline{\psfig{file=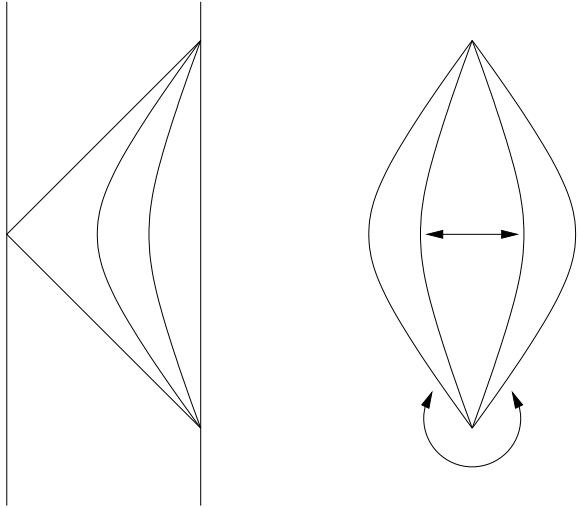,width=3.in}}
\caption{1. Anti-de Sitter space. Two horospheres and a horizon are
shown. The vertical lines represent timelike infinity. 
2. Causal structure of Randall-Sundrum model with compact fifth
dimension. The arrows denote identifications.}
\label{fig:global}
\end{figure}

Both models considered by RS use five dimensional
AdS. In horospherical coordinates the metric is
\be
\label{eqn:metric1}
 ds^2 = e^{-2y/l} \eta_{ij} dx^i dx^j + dy^2
\ee
where $\eta_{\mu\nu}$ is the four dimensional Minkowski metric and $l$
the AdS radius. The global structure of AdS is shown in figure
\ref{fig:global}. Horospherical coordinates break down at the
horizon $y=\infty$. 

In their first model \cite{RS1}, RS slice AdS along the horospheres
at $y=0$ and $y=y_c >0$, retain the portion $0<y<y_c$ and 
assume $Z_2$ reflection symmetry at each boundary plane. 
This gives a jump in extrinsic curvature at these
planes, yielding two domain walls of equal and opposite tension
\be
\label{eqn:tuning}
 \sigma = \pm\frac{6}{\kappa^2 l}
\ee
where $\kappa^2 = 8\pi G$ and $G$ is the five dimensional Newton constant.
The wall at $y=0$ has positive tension and the wall at $y=y_c$ has
negative tension. Mass scales on the negative tension wall are
exponentially suppressed relative to those on the positive tension
one. This provides a solution of the hierarchy problem provided we
live on the negative tension wall. The global structure is shown 
in figure \ref{fig:global}.

\begin{figure}
\centerline{\psfig{file=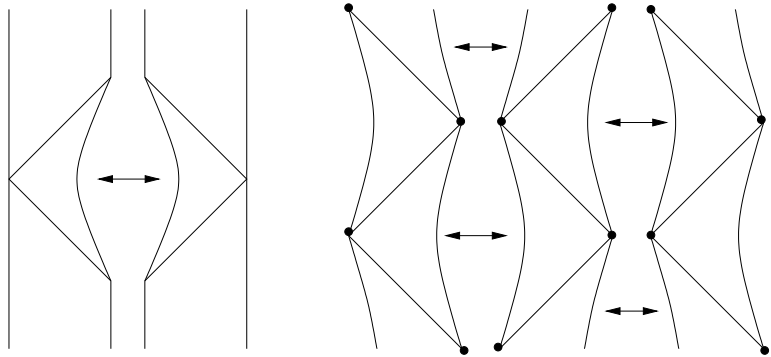,width=4.in}}
\caption{Possible causal structures for Randall-Sundrum model with
non-compact fifth dimension. The dots denote points at infinity.}
\label{fig:global2}
\end{figure}

The second RS model \cite{RS2} is obtained from the first by taking $y_c
\rightarrow \infty$. This makes the negative tension wall approach
the AdS horizon, which includes a point at infinity. RS say that their
model contains only one wall so presumably the idea is that the
negative tension brane is viewed as an auxiliary device to set up
boundary conditions. However, if the geometry
makes sense then it should be possible to discuss it without reference
to this limiting procedure involving negative tension objects. If one
simply slices AdS along a positive tension wall at $y=0$ and assumes
reflection symmetry then there are several ways to analytically
continue the solution across the horizon. These have been discussed in
\cite{gary,cvetic1,cvetic2,rep}. 
There are two obvious choices of continuation. The
first is simply to assume that beyond the horizon, the solution is
pure AdS with no domain walls present. This is shown in figure
\ref{fig:global2}. An alternative, which seems closer in spirit to the
geometry envisaged by RS, is to include further domain walls beyond
the horizon, as shown in figure \ref{fig:global2}. In this case, there
are infinitely many domain walls present. 

\sect{Black string in AdS}

Let us first rewrite the AdS metric \ref{eqn:metric1} by introducing
the coordinate $z=le^{y/l}$. The metric is then manifestly conformally
flat:
\be
 ds^2 = \frac{l^2}{z^2} (dz^2 + \eta_{ij}dx^i dx^j).
\ee
In these coordinates, the horizon lies at $z=\infty$ while the
timelike infinity of AdS is at $z=0$. We now note that if the Minkowski
metric within the brackets is replaced by {\it any} Ricci flat
metric then the Einstein equations (with negative cosmological
constant) are still satisfied\footnote{
This procedure was recently discussed for general p-brane solutions in
\cite{dom}.}. 
A natural choice for a metric describing a black hole on a domain wall
at fixed $z$ is to take this Ricci flat metric to be the Schwarzschild
solution:
\be
 ds^2 = \frac{l^2}{z^2}(-U(r)dt^2 + U(r)^{-1}dr^2 + r^2 (d\theta^2 +
\sin^2 \theta d\phi^2) + dz^2)
\ee
where $U(r) = 1-2M/r$. This metric describes a black string in AdS. 
Including a reflection symmetric
domain wall in this spacetime is trivial: surfaces of constant $z$
satisfy the Israel equations provided the domain wall tension
satisfies equation \ref{eqn:tuning}. For a domain wall at $z=z_0$,
introduce the coordinate $w=z-z_0$. The metric on both sides of the
wall can then be written
\be
 ds^2 =  \frac{l^2}{(|w|+z_0)^2}(-U(r)dt^2 + U(r)^{-1}dr^2 + r^2 (d\theta^2 +
\sin^2 \theta d\phi^2) + dw^2)
\ee
with $-\infty < w <\infty$ and the wall is at $w=0$. It would be
straightforward to use the same method to construct a black string
solution in the presence of a {\it thick} domain wall. 

The induced metric on a domain wall placed at $z=z_0$
can be brought to the standard Schwarzschild form by rescaling the
coordinates $t$ and $r$. The ADM mass as measured by an inhabitant of
the wall would be $M_*=Ml/z_0$. The proper radius of the
horizon in five dimensions is $2M_*$. 
The AdS length radius $l$ is required to
be within a few orders of magnitude of the Planck length \cite{RS2} so
black holes of astrophysical mass must have $M/z_0 \gg 1$. 
If one included a second domain wall with
negative tension then the ADM mass on that wall would be exponentially
suppressed relative to that on the positive tension wall.

Our solution has an Einstein metric so the Ricci scalar and square of
the Ricci tensor are finite everywhere. However the square of the
Riemann tensor is
\be
 R_{\mu\nu\rho\sigma} R^{\mu\nu\rho\sigma} = \frac{1}{l^4}\left(40 +
\frac{48 M^2 z^4}{r^6}\right),
\ee
which diverges at the AdS horizon $z=\infty$ as well as at the black
string singularity at $r=0$. We shall have more to say about this later.

It is important to examine the behaviour of geodesics in this
spacetime. Let $u$ denote the velocity along a timelike or null
geodesic with respect to an affine parameter $\lambda$ (taken to be
the proper time in the case of a timelike geodesic). 
The Killing vectors $k=\partial/\partial t$ and
$m=\partial/\partial \phi$ give rise to the conserved quantities
$E=-k\cdot u$ and $L=m\cdot u$. Rearranging these gives
\be
 \frac{dt}{d\lambda} = \frac{E z^2}{U(r) l^2}
\ee
and
\be
 \frac{d\phi}{d\lambda} = \frac{L z^2}{r^2 l^2},
\ee
for motion in the equatorial plane ($\theta \equiv \pi/2$). The
equation describing motion in the $z$-direction is simply
\be
 \frac{d}{d\lambda}\left(\frac{1}{z^2}\frac{dz}{d\lambda}\right) =
 \frac{\sigma}{z l^2}, 
\ee
where $\sigma=0$ for null geodesics and $\sigma=1$ for timelike geodesics.
The solutions for null geodesics are $z=\mathrm{constant}$ or 
\be
 z = -\frac{z_1 l}{\lambda},
\ee
The solution for timelike geodesics is
\be
 z = -z_1 \mathrm{cosec} (\lambda/l).
\ee
In both cases, $z_1$ is a constant and we have shifted $\lambda$ 
so that $z\rightarrow \infty$ as $\lambda \rightarrow 0-$.
The (null) solution $z=\mathrm{const}$ is simply a 
null geodesic of the four dimensional Schwarzschild solution. We are
more interested in the other solutions because they appear to reach the 
singularity at $z=\infty$. The radial motion is given by
\be
 \left(\frac{dr}{d\lambda}\right)^2 +
\frac{z^4}{l^4}\left[\left(\frac{l^2}{z_1^2} + \frac{L^2}{r^2}\right) U(r) -
E^2 \right] = 0.
\ee
Now introduce a new parameter $\nu=-z_1^2/\lambda$ for null geodesics
and $\nu=-(z_1^2/l)\cot (\lambda/l)$ for timelike geodesics. We also define
new coordinates
$\tilde{r} = z_1 r/l$, $\tilde{t} = z_1 t/l$, and new constants
$\tilde{E} = z_1 E/l$, $\tilde{L} = z_1^2 L/l^2$ and $\tilde{M} = z_1
M/l$. The radial equation becomes
\be
 \left(\frac{d\tilde{r}}{d\nu}\right)^2 + \left(1 + \frac{\tilde{L}^2}
{\tilde{r}^2}\right)\left(1-\frac{2\tilde{M}}{\tilde{r}}\right) =
\tilde{E}^2,
\ee
which is the radial equation for a {\it timelike} geodesic in a four
dimensional Schwarzschild solution of mass $\tilde{M}$
\cite{wald}. (This is the ADM mass for an observer with $z=z_0=l^2/z_1$.)
Note that $\nu$ is the proper time along such a
geodesic. It should not be surprising that a null geodesic in five
dimensions is equivalent to a timelike geodesic in four dimensions:
the non-trivial motion in the fifth dimension gives rise to a mass in
four dimensions. What is perhaps surprising is the relationship
between the four and five dimensional affine parameters $\nu$ and
$\lambda$.

We are interested in the behaviour near the singularity, i.e. as
$\lambda \rightarrow 0-$. This is equivalent to $\nu \rightarrow \infty$
i.e. we need to study the late time behaviour of four dimensional
timelike geodesics. If such geodesics hit the {\it Schwarzschild}
singularity at $\tilde{r}=0$ then they do so at finite $\nu$. For
infinite $\nu$ there are two possibilities \cite{wald}. The first is
that the geodesic reaches $\tilde{r}=\infty$. The second can occur 
only if $\tilde{L}^2 > 12\tilde{M}^2$, when it is possible to have 
bound states (i.e. orbits restricted to a finite range of $\tilde{r}$)
outside the Schwarzschild horizon.

The orbits that reach $\tilde{r}=\infty$ have late time behaviour
$\tilde{r} \sim \nu\sqrt{\tilde{E}^2-1}$ and hence
\be
 r \sim -\frac{z_1 l}{\lambda} \sqrt{\tilde{E}^2-1}
\ee
as $\lambda \rightarrow 0-$. Along such geodesics, the squared Riemann
tensor does {\it not} diverge.
The bound state geodesics behave differently. These
remain at finite $r$ and therefore the square of the Riemann
tensor {\it does} diverge
as $\lambda \rightarrow 0-$. They orbit the black string
infinitely many times before hitting the singularity, but do so in
finite affine parameter.

\bigskip

It appears that some geodesics encounter a curvature singularity at
the AdS horizon whereas others might not because scalar curvature
invariants do not diverge along them. It is possible that only part
of the surface $z=\infty$ is singular. To decide whether or not
this is true, we turn to a calculation of the Riemann tensor in an 
orthonormal frame parallelly propagated along a timelike geodesic that
reaches $z=\infty$ but for which the squared Riemann tensor does not
diverge (i.e. a non-bound state geodesics). 
The tangent vector to such a geodesic (with $L=0$)
can be written
\be
 u^{\mu} = \left(\frac{z}{l}\sqrt{\frac{z^2}{z_1^2}-1}, \frac{E z^2}{U(r)
l^2}, \frac{z^2}{l^2}\sqrt{E^2-\frac{l^2}{z_1^2} U(r)},0 ,0\right),
\ee
where we have written the components in the order
$(z,t,r,\theta,\phi)$. A unit normal to the geodesic is
\be
 n^{\mu} = \left(0, -\frac{z z_1}{l^2 U(r)}\sqrt{E^2 -\frac{l^2}{z_1^2}
U(r)}, -\frac{E z_1 z}{l^2},0,0\right).
\ee
It is straightforward to check that this is parallelly propagated along
the geodesic i.e. $u\cdot\nabla n^\mu =0$. 
These two unit vectors can be supplemented by three other parallelly
propagated vectors to form an orthonormal set. However the divergence
can be exhibited using just these two vectors. One of the curvature
components in this parallelly propagated frame is
\be
 R_{(u)(n)(u)(n)} \equiv R_{\mu\nu\rho\sigma} u^{\mu} n^{\nu} u^{\rho}
n^{\sigma} = \frac{1}{l^2}\left(1 -\frac{2 M z^4}{z_1^2 r^3}\right),
\ee
which diverges along the geodesic as $\lambda \rightarrow 0$. The
black string solution therefore has a curvature singularity at the AdS
horizon.

\bigskip

It is well known that black string solutions in asymptotically flat
space are unstable to long wavelength perturbations
\cite{greg}. A black hole is entropically preferred to a sufficiently
long segment of string. The string's horizon therefore has a tendency
to ``pinch off'' and form a line of black holes. 
One might think that a similar instability occurs for
our solution. However, AdS acts like a confining box which
prevents fluctuations with wavelengths much greater than
$l$ from developing. If an instability occurs then it must do so at
smaller wavelengths.

If the radius of curvature of the string's horizon is 
sufficiently small then the AdS curvature will be negligible there and
the string will behave as if it were in asymptotically flat
space. This means that it will be unstable to perturbations with
wavelengths of the order of the horizon radius $2M_*=2Ml/z$. At
sufficiently large $z$, such perturbations will fit into the AdS box,
i.e. $2M_* \ll l$, so an instability can occur near the AdS horizon. 
However for $M/z \gg 1$, the potential instability 
occurs at wavelengths much greater than $l$ and is
therefore not possible in AdS. Therefore the black string
solution is unstable near the AdS horizon but stable far from it. 

We conclude that, near the AdS horizon, the black string has a tendency
to ``pinch off'' but further away it is stable. After pinching off,
the string becomes a stable ``black cigar''
which would extend to infinity in AdS if the domain wall were not
present, but not to the AdS horizon.
The cigar's horizon acts as if it has a tension which balances the force
pulling it towards the centre of AdS. We showed above that if our
domain wall is at $z=z_0$ then a black hole of astrophysical mass has
$M/z_0 \gg 1$, corresponding to the part of the black
cigar far from the AdS horizon. Here, the metric 
will be well approximated by the black string metric 
so the induced metric on the wall will be Schwarzschild and the 
predictions of four dimensional general relativity will be recovered. 

\sect{Discussion}

Any phenomenologically successful theory in which our universe is
viewed as a brane must reproduce the large-scale 
predictions of general relativity on the brane. 
This implies that gravitational collapse of uncharged non-rotating 
matter trapped on the brane ultimately settles down to a steady 
state in which the induced metric on the brane is Schwarzschild. 
In the higher dimensional theory, such a solution could be
a localized black hole or an extended object intersecting the brane. 
We have investigated these alternatives in the models proposed by 
Randall and Sundrum (RS). The obvious choice of five dimensional
solution in the first case is Schwarzschild-AdS. However
we have shown (in the Appendix) that it is not possible to intersect 
this with a vacuum domain wall so it cannot be the 
final state of gravitational collapse on the wall. 

We have presented a solution that describes a black string in AdS. It
{\it is} possible to intersect this solution with a vacuum domain wall
and the induced metric is Schwarzschild. The solution can therefore
be interpreted as a black hole on the wall. The AdS horizon
is singular. Scalar curvature invariants only diverge if this 
singularity is approached 
along the axis of the string. However, curvature components diverge in a 
frame parallelly propagated along any timelike geodesic that reaches
the horizon. This singularity can be removed if we use the first 
RS model in which there are two domain walls present 
and we live on a negative tension
wall. However if we wish to use the second RS model (with a
non-compact fifth dimension) then the singularity will be visible from
our domain wall. In \cite{ppwave}, it
was argued that anything emerging from a singularity at the AdS
horizon would be heavily red-shifted before reaching us 
and that this might ensure that
physics on the wall remains predictable in spite of the
singularity. However we regard singularities as a pathology of the
theory since, in principle, arbitrarily large fluctuations can emerge 
from the singularity and the red-shift is finite.

Fortunately, it turns out that our solution is unstable near the AdS
horizon. We have suggested that it will decay to a stable
configuration resembling a cigar that extends out to infinity in AdS
but does not reach the AdS horizon. The solution becomes finite in extent 
when the gravitational effect of the domain wall is included.
Our domain wall is situated far
from the AdS horizon so the induced metric on the wall will be very nearly 
Schwarzschild. Since the cigar does not extend as far as the AdS horizon,
it does not seem likely that there will be a singularity there. 
Similar behaviour was recently found in a non-linear treatment 
of the RS model \cite{ppwave}. 
It was shown that pp-waves corresponding to Kaluza-Klein modes
are singular at the AdS horizon. These pp-waves are not localized to
the domain wall. The only pp-waves
regular at the horizon are the ones corresponding to gravitons
confined to the wall. We suspect that perturbations of the flat
horospheres of AdS that do not vanish near the horizon will generically
give rise to a singularity there.

It seems likely that there are other solutions that give rise to the
Schwarzschild solution on the domain wall. For example, the metric
outside a star on the wall would be Schwarzschild. If the cigar solution
was the only stable solution giving Schwarzschild on the wall then it would
have to be possible to intersect it with a non-vacuum domain wall
describing such a star. However, it is then not possible to choose the
equation of state for the matter on the wall, for reasons 
analogous to those discussed in the Appendix. Our solution
is therefore not capable of describing generic stars. If this is the
case then one might wonder whether there are other solutions describing
black holes on the wall. We conjecture that the cigar solution is the
unique stable vacuum solution with a regular AdS horizon that describes a
non-rotating uncharged black hole on the domain wall.

\begin{center}
{\bf Acknowledgments}
\end{center}
We have benefitted from discussions with D. Brecher and G. Gibbons. AC was
supported by Pembroke College, Cambridge.

\appendix
\renewcommand{\theequation}{\arabic{equation}}
\section*{Appendix}
\setcounter{equation}{0}

One candidate for a black hole formed by gravitational collapse
on a domain wall in AdS is the Schwarzschild-AdS solution, which has metric
\be
 ds^2 = -U(r)dt^2 + U(r)^{-1}dr^2 + r^2(d\chi^2+\sin^2 \chi d\Omega^2),
\ee
where $d\Omega^2$ is the line element on a unit 2-sphere and 
\be
 U(r) = 1-\frac{2M}{r^2} + \frac{r^2}{l^2}.
\ee
The parameter $M$ is related to the mass of the black hole. We have
not yet included the gravitational effect of the wall. We shall focus
on the second RS model so we want a single positive tension domain
wall with the spacetime reflection symmetric in the wall. Denote the
spacetime on the two sides of the wall as $(+)$ and $(-)$. Let $n$ be
a unit (spacelike) normal to the wall pointing out of the $(+)$
region. The tensor $h_{\mu\nu} = g_{\mu\nu} - n_{\mu} n_{\nu}$ 
projects vectors onto
the wall, and its tangential components give the induced metric on
the wall. The extrinsic curvature of the wall is defined by
\be
 K_{\mu\nu} = h_{\mu}^{\rho} h_{\nu}^{\sigma} \nabla_{\rho} n_{\sigma}
\ee
and its trace is $K=h^{\mu\nu}K_{\mu\nu}$. The energy momentum 
tensor $t_{\mu\nu}$
of the wall is given by varying its action with respect to the induced
metric. The gravitational effect of the
domain wall is given by the Israel junction conditions \cite{israel},
which relate the discontinuity in the extrinsic curvature at the wall
to its energy momentum:
\be
 [K_{\mu\nu} - Kh_{\mu\nu}]^+_- = \kappa^2 t_{\mu\nu}
\ee 
(see \cite{dddw} for a simple derivation of this equation). Here
$\kappa^2 = 8\pi G$ where $G$ is the five dimensional Newton constant.
This can be rearranged using reflection symmetry to give
\be
 K_{\mu\nu} = \frac{\kappa^2}{2}\left(t_{\mu\nu} -
\frac{t}{3}h_{\mu\nu}\right), 
\ee
where $t = h^{\mu\nu}t_{\mu\nu}$. 

Cylindrical symmetry dictates that we should consider a domain wall 
with position given by $\chi = \chi(r)$. The unit normal to the $(+)$ side
can be written
\be
 n = \frac{\epsilon r}{\sqrt{1+Ur^2{\chi'}^2}} \left(d\chi - \chi'
dr\right),
\ee
where $\epsilon = \pm 1$ and a prime denotes a derivative with respect
to $r$. The timelike tangent to the wall is
\be
 u = U^{-1/2} \frac{\partial}{\partial t},
\ee
and the spacelike tangents are
\be
 t = \sqrt{\frac{U}{1+U r^2 {\chi}^2}} \left(\chi'
\frac{\partial}{\partial \chi} + \frac{\partial}{\partial r}\right),
\ee
\be
 e_{\theta} = \frac{1}{r \sin \chi} \frac{\partial}{\partial \theta},
\ee
\be
 e_{\phi} = \frac{1}{r \sin \chi \sin \theta}
\frac{\partial}{\partial\phi}.
\ee
The non-vanishing components of the extrinsic curvature in this basis are
\be
 K_{uu} = \frac{\epsilon U' r \chi'}{2\sqrt{1+U r^2{\chi'}^2}},
\ee
\be
 K_{\theta\theta} = K_{\phi\phi} = \frac{\epsilon}{\sqrt{1+U
r^2{\chi'}^2}} \left(\frac{\cot \chi}{r} - U \chi'\right),
\ee
\be
 K_{tt} = -\frac{\epsilon}{\left(1+Ur^2{\chi'}^2\right)^{3/2}} \left(
{\chi'}^3 U^2 r^2 + 2\chi' U + U r \chi'' + U' r \chi'/2\right).
\ee

A vacuum domain wall has
\be
 t_{\mu\nu} = -\sigma h_{\mu\nu},
\ee
where $\sigma$ is the wall's tension. The Israel conditions are
\be
 K_{\mu\nu} = \frac{\kappa^2}{6}\sigma h_{\mu\nu}.
\ee
These reduce to
\be
 -K_{uu}=K_{tt}=K_{\theta\theta}=\frac{\kappa^2}{6}\sigma.
\ee
It is straightforward to verify that these equations have no
solution. A solution {\it can} be found for a non-vacuum domain wall
with energy-momentum tensor
\be
 t_{\mu\nu} = \mathrm{diag}(\sigma,p,p,p,0),
\ee
since then we have three unknown functions ($\sigma(r),p(r),\chi(r)$)
and three equations. However this does not allow an equation of state
to be specified in advance. We are only interested in {\it vacuum}
solutions since these describe the final state of
gravitational collapse on the brane.

\end{document}